\documentclass[12pt]{aastex}

\newcommand{\scale}[3]{\left(\frac{#1}{#2}\right)^{#3}}
\newcommand{\sub}[1]{_{\rm #1}}
\renewcommand{\vec}[1]{\mbox{\boldmath $#1$}}

\shorttitle{Formation of Terrestrial Planets}
\shortauthors{Kokubo \& Genda}

\begin{document}

\title{Formation of Terrestrial Planets from Protoplanets under a Realistic
Accretion Condition}

\author{Eiichiro Kokubo}
\affil{Division of Theoretical Astronomy, National Astronomical
 Observatory of Japan, Osawa, Mitaka, Tokyo, 181-8588, Japan}
\email{kokubo@th.nao.ac.jp}

\and

\author{Hidenori Genda}
\affil{Department of Earth and Planetary Sciences, Tokyo Institute of
 Technology, Ookayama, Meguro, Tokyo, 152-8551, Japan}
\email{genda@geo.titech.ac.jp}

\begin{abstract}

The final stage of terrestrial planet formation is known as the giant
 impact stage where protoplanets collide with one another to form
 planets.    
So far this stage has been mainly investigated by $N$-body simulations
 with an assumption of perfect accretion in which all collisions lead to 
 accretion. 
However, this assumption breaks for collisions with high velocity and/or
 a large impact parameter.  
We derive an accretion condition for protoplanet collisions
 in terms of impact velocity and angle and masses of colliding bodies,
 from the results of numerical collision experiments. 
For the first time, we adopt this realistic accretion condition in
 $N$-body simulations of terrestrial planet formation from protoplanets.  
We compare the results with those with perfect accretion and show how
 the accretion condition affects terrestrial planet formation.    
We find that in the realistic accretion model, about half of
 collisions do not lead to accretion. 
However, the final number, mass, orbital elements, and even growth
 timescale of planets are barely affected by the accretion condition.  
For the standard protoplanetary disk model, typically two Earth-sized
 planets form in the terrestrial planet region over about $10^8$ years in
 both realistic and perfect accretion models.
We also find that for the realistic accretion model, the spin
 angular velocity is about 30\% smaller than that for the perfect
 accretion model that is as large as the critical spin angular velocity
 for rotational instability.
The spin angular velocity and obliquity obey Gaussian and isotropic
 distributions, respectively, independently of the accretion condition.

\end{abstract}

\keywords{planets and satellites: formation---methods: numerical}

\section{INTRODUCTION}

It is generally accepted that the final stage of terrestrial planet
 formation is the giant impact stage where protoplanets or planetary
 embryos formed by oligarchic growth collide with one another to form
 planets \citep[e.g.,][]{w85,ki98}.
This stage has been mainly studied by $N$-body simulations.
So far all $N$-body simulations have assumed perfect accretion in which
 all collisions lead to accretion \citep[e.g.,][]{acl99,c01}.
However, this assumption would be inappropriate for grazing impacts that  
 may result in escape of an impactor or hit-and-run.
By performing Smoothed-Particle Hydrodynamic (SPH) collision simulations,
 \cite{aa04} estimated that more than half of all collisions between 
 like-sized protoplanets do not simply result in accumulation of a larger
 protoplanet, and this inefficiency lengthens the timescale of planet
 formation by a factor of 2 or more, relative to the perfect accretion
 case.     
The accretion inefficiency can also change planetary spin.
\cite{ki07} found that under the assumption of perfect accretion, the
 typical spin angular velocity of planets is as large as the critical
 spin angular velocity for rotational instability.
However, in reality, the grazing collisions that have high angular
 momentum are likely to result in a hit-and-run, while nearly head-on
 collisions that have small angular momentum lead to accretion. 
In other words, small angular momentum collisions are selective in
 accretion.  
Thus, the accretion inefficiency may lead to slower planetary spin,
 compared with the perfect accretion case.

The goal of this paper is to clarify the statistical properties of
 terrestrial planets formed by giant impacts among protoplanets under a
 realistic accretion condition.  
We derive an accretion condition for protoplanet collisions in terms of
 collision parameters, masses of colliding protoplanets and impact
 velocity and angle, by performing collision experiments with an SPH
 method.   
We implement the realistic accretion condition in $N$-body simulations
 and probe its effect to further generalize the model of terrestrial
 planet formation.   
We derive the statistical dynamical properties of terrestrial planets
 from results of a number of $N$-body simulations and compare the
 results with those in \cite{kki06} and \cite{ki07} where perfect
 accretion is adopted.  

In section 2, we outline the initial conditions of protoplanets and the
 realistic accretion condition.
Section 3 presents our results, where we show the statistics of
 collision parameters and basic dynamical properties of planets. 
Section 4 is devoted to a summary and discussions.

\section{METHOD OF CALCULATION}
\label{sec:method}

\subsection{Initial Conditions}

We perform $N$-body simulations of terrestrial planet formation
 starting from protoplanets.
We consider gas-free cases without giant planets residing outside the
 terrestrial planet region to clarify the basic dynamics. 
To compare with the results with perfect accretion, we adopt the same
 protoplanet system as those in \cite{kki06} and \cite{ki07}, which is
 formed by oligarchic growth from a planetesimal disk whose surface
 density distribution is given by 
\begin{equation}
 \Sigma = 10\left(\frac{r}{\rm 1AU}\right)^{-3/2}{\rm g cm}^{-2},
\end{equation}
 with inner and outer edges, $r\sub{in}=0.5$AU and $r\sub{out}=1.5$AU,
 where $r$ is the radial distance from the central star.
This disk model is the standard disk model for solar system formation
 and 50\% more massive than the minimum-mass disk \citep{h81}.
In the oligarchic growth model, the mass of a protoplanet $M$ is given
 by the isolation mass
\begin{equation} 
 \label{eq:m_iso}
 M\sub{iso}
 \simeq 2\pi ab\Sigma = 
 0.16
 \scale{\tilde{b}}{10}{3/2}
 \scale{a}{1{\rm AU}}{3/4}M_\oplus,
\end{equation}
 where $a$ is the semimajor axis, $b$ is the orbital separation between
 adjacent protoplanets,
 $\tilde{b}=b/r\sub{H}$, $r\sub{H}$ is the Hill radius
 $r\sub{H}=(2M\sub{iso}/3M_\odot)^{1/3}a$ of the protoplanet,
 $M_\odot$ is the mass of the central star, and
 $M_\oplus$ is Earth mass \citep{ki00}.  
We set the orbital separation of protoplanets as $\tilde{b}=10$ that
 is the typical value in $N$-body simulations \citep[e.g.,][]{ki00, ki02}. 
The initial eccentricities $e$ and inclinations $i$ of protoplanets are
 given by the Rayleigh distribution with dispersions $\langle
 e^2\rangle^{1/2}=2\langle i^2\rangle^{1/2}=0.01$
 (the unit of $i$ is radian) \citep{im92}. 
We set the bulk density of protoplanets as $\rho=3$ gcm$^{-3}$.
The initial protoplanet system has the number of protoplanets $n=16$ in 
 $r\sub{in}\leq a \leq r\sub{out}$,
 total mass $M\sub{tot}\simeq 2.3M_\oplus$, specific angular momentum 
 $j \simeq 0.95j_\oplus$, and mean semimajor axis
 $\bar{a}=j^2/GM_\odot=0.91$AU, where $j_\oplus=\sqrt{GM_\odot
 a_\oplus}$ and $a_\oplus$ is the semimajor axis of Earth. 
For each accretion model, we perform 50 runs with different initial angular
 distributions of protoplanets.

\subsection{Orbital Integration}

The orbits of protoplanets are calculated by numerically integrating
 the equations of motion of protoplanets.
We set the mass of the central star equal to solar mass.
For numerical integration, we use the modified Hermite scheme for
 planetary $N$-body simulation \citep{kym98,km04} with the hierarchical
 timestep \citep{m91}. 
For the calculation of mutual gravity among protoplanets, we use the
 Phantom GRAPE scheme \citep{nmh06}. 
The simulations follow the evolution of protoplanet systems for 
 $3\times 10^8$ years until only a few planets remain.

\subsection{Accretion Condition}

\begin{figure}
\epsscale{0.5}
\plotone{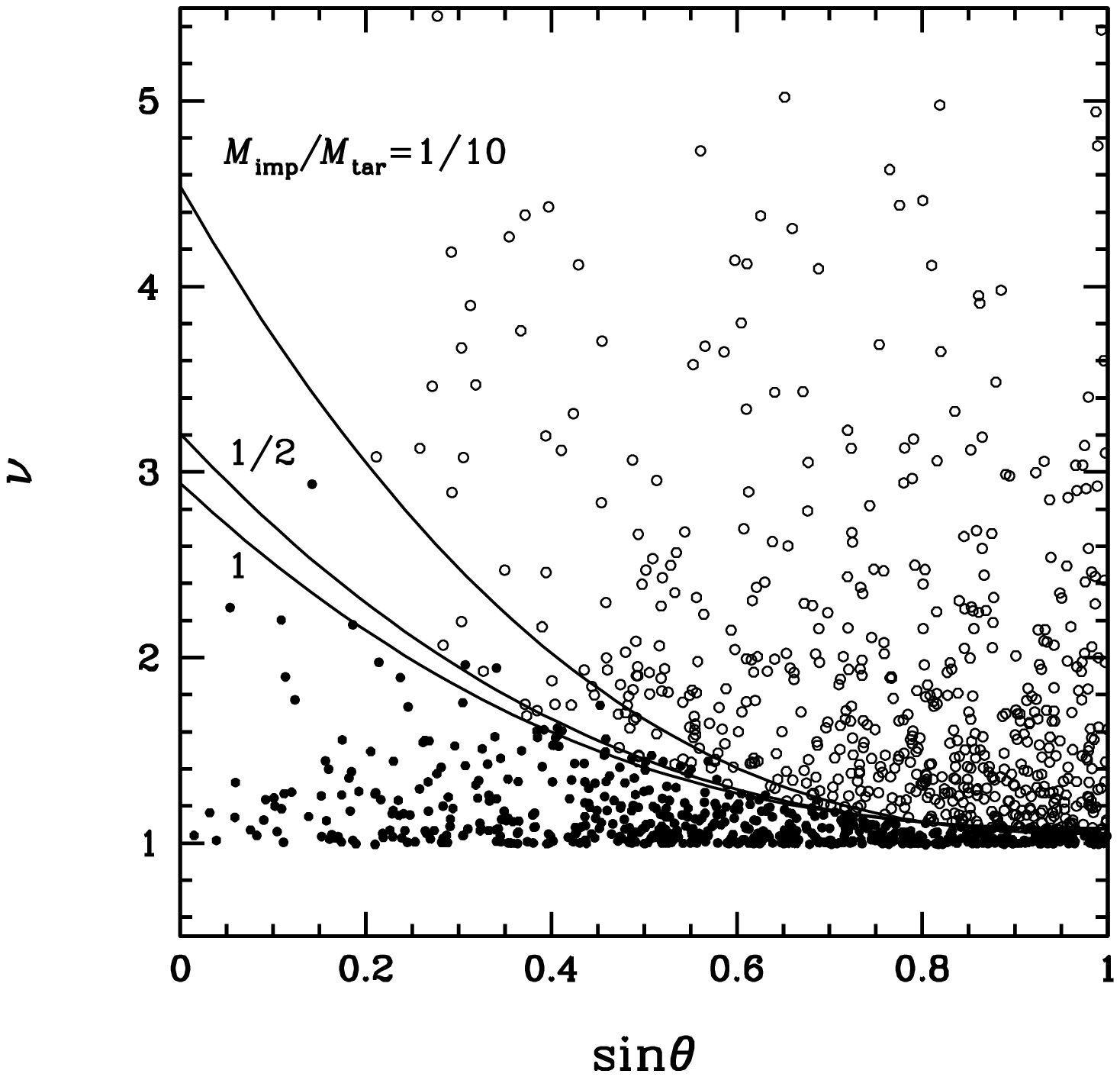}
\caption{
\label{fig:sintheta-nu}
Scaled impact velocity $\nu$ is plotted against the scaled impact
 parameter $\sin\theta$ for accretionary (filled circles) and
 hit-and-run (open circles) collisions in 50 runs of the realistic
 accretion model.
The solid curves show the scaled critical impact velocity
 $v\sub{cr}/v\sub{esc}$ for $M\sub{imp}/M\sub{tar}=1/10$, 1/2, and 1. 
}
\epsscale{1.0}
\end{figure}

During orbital integration, when two protoplanets contact, a collision
 occurs. 
We define an impactor/target as a smaller/larger one of two colliding
 bodies.
We obtained a realistic accretion condition of protoplanets by
 performing SPH collision simulations. 
The standard SPH method \citep{m92,c04} was used with the Tillotson
 equation of state \citep{m89}.
We assumed differentiated protoplanets with 30\% core and 70\% mantle in
 mass.
Protoplanets were represented by 20,000 particles in most runs and
 60,000 particles in high-resolution runs.
We systematically varied the mass ratio of the impactor and target as
 $M\sub{imp}/M\sub{tar} = 1$, 2/3, 1/2, 1/3, 1/4, 1/6, and 1/9, their
 total mass as $M\sub{col} = M\sub{imp} + M\sub{tar} = 0.2$-$2
 M_\oplus$, and the impact velocity and angle as  
 $v\sub{imp} = 1.0$-$3.0 v\sub{esc}$ at 0.02 or $0.2 v\sub{esc}$
 intervals and $\theta = 0^\circ$-$75^\circ$ at $15^\circ$ intervals,
 where $v\sub{esc}$ is the mutual surface escape velocity 
 $v\sub{esc} = [2GM\sub{col}/(R\sub{imp}+R\sub{tar})]^{1/2}$,
 $R\sub{imp}$ and $R\sub{tar}$ are the radii of the impactor and
 target, and $\theta$ is the angle between the surface normal and the impact
 trajectory ($\theta = 0^\circ$ for a head-on collision and $\theta =
 90^\circ$ for a grazing encounter).  
Based on the results of SPH collision simulations, we derived an
 empirical formula for the critical impact velocity, below 
 which a collision leads to accretion, in terms of the masses of the
 impactor and target and the impact angle as 
\begin{equation}
 \frac{v\sub{cr}}{v\sub{esc}} =
 c_1\left(\frac{M\sub{tar}-M\sub{imp}}{M\sub{col}}\right)^2(1-\sin\theta)^{5/2}  
+ c_2 \left(\frac{M\sub{tar}-M\sub{imp}}{M\sub{col}}\right)^2
+ c_3 (1-\sin\theta)^{5/2} + c_4,
\end{equation}
 where $c_1 = 2.43$, $c_2 = -0.0408$, $c_3 = 1.86$, and $c_4 = 1.08$ are
 numerical constants. 
The critical impact velocity is a decrease function of $\sin\theta$ as
 seen in Figure~\ref{fig:sintheta-nu}.
For collisions of equal-mass protoplanets, this condition agrees well
 with the results of SPH collision simulations by \cite{aa04}.  
The reason for this agreement is that we performed almost the same
 collision simulations by essentially the same method, which confirms
 the robustness of the results. 
The details of the collision simulations and the accretion condition
 will be presented in a separate paper. 
In the realistic accretion model we use this accretion condition.
In accretion, the position and velocity of the center of mass and
 the angular momentum are conserved.

If the impact velocity $v = |\vec{v}| = \sqrt{v\sub{n}^2 + v\sub{t}^2}$
 is higher than $v\sub{cr}$, two colliding bodies 
 are bounced with the rebound velocity $\vec{v}'$,  
 $v'\sub{n} = 0$ and $v'\sub{t} = \max(v\sub{t}, v\sub{esc})$, 
 where subscripts n and t mean normal and tangential components of the
 velocity, respectively.
The change in spin angular momentum brought about by hit-and-run
 collisions is not taken into account since it is usually much smaller
 than that caused by accretionary collisions.
These prescriptions follow the results of the collision experiments.

We assume that, initially, protoplanets have no spin angular momenta.
We track the spin angular momentum of planets resulting from accretion.
When accretion occurs, the orbital angular momenta of two colliding
 bodies about their center of mass and their spin angular momenta are
 added to the spin angular momentum of a merged body.
The spin angular velocity of planets is computed by assuming a sphere of
 uniform density, in other words, the moment of inertia $I=(2/5)MR^2$,
 where $M$ and $R$ are the planetary mass and radius, respectively.


\section{RESULTS}
\label{sec:result}

We compare the results of the realistic accretion model to those
 of the perfect accretion model \citep{kki06,ki07}.
The statistics of collision parameters and basic dynamical properties of
 planets derived from 50 runs for each accretion model are summarized in 
 Tables~\ref{tab:number} and \ref{tab:spin}.

\subsection{Statistics of Collision Parameters}
\label{sec:collision_parameter}


We record the collision parameters in all runs.
The numbers of hit-and-run $n\sub{har}$ and accretionary $n\sub{acc}$
 collisions are summarized in Table~\ref{tab:number}. 
In the realistic accretion model, the total number of collisions in 50
 runs is 1211.
In a run, the average numbers of hit-and-run and accretionary collisions
 are $\langle n\sub{har}\rangle=11.8$ and $\langle n\sub{acc}\rangle=12.4$, 
 respectively, which means on average 49\% of collisions results in
 hit-and-run.   
Note that the average number of accretionary collisions for the realistic
 accretion model is almost the same as that for the perfect accretion model. 
This suggests that the growth timescale of planets is independent of the
 accretion model.  
We will discuss this later.

On the giant impact stage, the last one or a few
 large collisions determine the final spin angular momentum of planets
 \citep{acl99, ki07}.
In this case, the spin angular velocity is estimated as
\begin{equation}
\label{eq:w}
 \omega \simeq \frac{5}{\sqrt{2}}
 \langle g(\gamma)^2\rangle^{1/2}
 \langle \sin\theta^2\rangle^{1/2}
 \langle \nu^2\rangle^{1/2} \omega\sub{cr},
\end{equation}
 where 
 $\gamma=M\sub{imp}/M\sub{col}$, 
 $g(\gamma)=\gamma(1-\gamma)\left[\gamma^{1/3}+(1-\gamma)^{1/3}\right]^{1/2}$, 
 and $\nu=v/v\sub{esc}$.
The critical spin angular velocity for rotational instability
 $\omega\sub{cr}$ is given by
\begin{equation}
 \label{eq:w_cr}
 \omega\sub{cr} = \left(\frac{GM}{R^3}\right)^{1/2} =
  3.3\left(\frac{\rho}{3{\rm gcm}^{-3}}\right)^{1/2} {\rm hr}^{-1},
\end{equation}
 where $\rho$ is the material density. 

In Figure~\ref{fig:sintheta-nu}, we plot the scaled impact velocity $\nu$
 against $\sin\theta$ for all collisions in all runs for the realistic
 accretion model.
We show that only collisions with small $\nu$ and/or $\sin\theta$ can
 lead to accretion due to the realistic accretion condition. 
For collisions forming Earth-sized planets ($M > M_\oplus/2$), 
 the RMS values of the collision parameters are 
 $\langle g(\gamma)^2\rangle^{1/2}=0.24$,
 $\langle \sin^2\theta\rangle^{1/2}=0.62$, and 
 $\langle \nu^2\rangle^{1/2}=1.19$ for the realistic accretion model,
 and 
 $\langle g(\gamma)^2\rangle^{1/2}=0.25$,
 $\langle \sin^2\theta\rangle^{1/2}=0.71$, and 
 $\langle \nu^2\rangle^{1/2}=1.40$ for the perfect accretion model.
The RMS impact parameter $\langle \sin^2\theta\rangle^{1/2}$ for the
 realistic accretion model is smaller than that for the perfect
 accretion model since small impact parameters are
 selective in the realistic accretion model. 
Also, for the same reason, the RMS impact velocity $\nu$ for the
 realistic accretion model is slightly smaller than that for the perfect
 accretion model.  
Using these RMS values, we obtain the typical spin angular velocity
 resulting from an accretionary collision as 2.0 hr$^{-1}$ and 2.8
 hr$^{-1}$ for the realistic and perfect accretion models,
 respectively.

\subsection{Statistics of Basic Dynamical Properties}


The average values with standard deviations for basic dynamical
 properties of planets: the number of planets,   
 $n$, the number of Earth-sized planets with $M>M_\oplus/2$,
 $n_M$, the number of planets in $r\sub{in}\leq a \leq r\sub{out}$,
 $n_a$, the in-situ accretion efficiency $f_a=M_a/M\sub{tot}$ where
 $M_a$ is the total mass of planets in $r\sub{in}\leq a \leq r\sub{out}$,
 and the growth timescale $T\sub{grow}$ defined as duration of accretion
 for each planet that undergoes at least two accretionary collisions
 are summarized in Table~\ref{tab:number}.  
The mass and orbital elements (semimajor axis $a$, eccentricity $e$, and
 inclination $i$) of the largest and second-largest planets are
 summarized in Table~\ref{tab:number}. 

First, we find no substantial differences between the results of the two
 accretion models. 
In both models, a typical resultant system consists of two Earth-sized
 planets and one or two smaller planets that are as large as the
 initial protoplanets \citep{kki06}.  
The two Earth-sized planets tend to form inside the initial distribution
 of protoplanets.
The resultant system for the realistic accretion model tends to be
 slightly wider than that for the perfect accretion model, which can be
 seen as having slightly larger $\langle n \rangle$ and smaller 
 $\langle n_a \rangle$ and $\langle f_a \rangle$.
This is because the realistic accretion model experiences more close
 encounters between protoplanets that diffuse the system.
It should be noted that though about half of collisions do not lead to
 accretion, this accretion inefficiency does not lengthen the growth
 timescale by a factor of two or more, relative to the perfect
 accretion model as expected by \cite{aa04}. 
The growth timescale for the realistic accretion model is only slightly
 longer than that for the perfect accretion model.
This is because even though collisions do not lead to accretion,
 the colliding bodies stay on the colliding orbits after the collision
 and thus the system is unstable and the next collision takes place
 shortly. 
In fact, the mean accretionary collision time for the realistic
 accretion model is $T\sub{col} = 11.1\pm 29.3\times 10^6$ years, which
 is as long as the mean collision time for the perfect accretion model 
 $T\sub{col}= 10.3\pm 30.3\times 10^6$ years.  
On the other hand, in the realistic accretion model, the mean collision
 time after a hit-and-run collision is short as
 $T\sub{col} = 3.6\pm 9.1\times 10^6$ years, where 38\% of collisions is 
 for the same pair and its mean collision time is even shorter as
 $T\sub{col} = 2.1\pm 5.6\times 10^6$ years.  

\begin{figure}
\epsscale{0.5}
\plotone{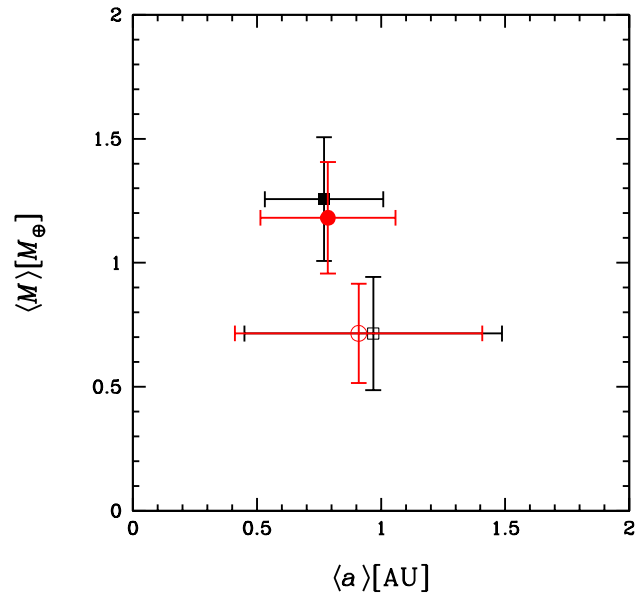}
\caption{
 \label{fig:a-m}
 Average semimajor axes and masses of the largest (filled symbols)
 and second-largest (open symbols) planets for realistic (circle) and
 perfect (square) accretion models.
 The error bars indicate 1-$\sigma$.
 }
\epsscale{1.0}
\end{figure}

Figure~\ref{fig:a-m} shows the average masses of the largest and
 second-largest planets against their average semimajor axes.
We find that there are no differences in either mass or semimajor axis
 of the planets for the realistic and perfect accretion models.
The largest planets with $M\simeq 1.2 M_\oplus$ tend to form around
 $\langle a_1\rangle\simeq 0.8$ AU, while the second-largest planets
 with $M\simeq 0.7 M_\oplus$ is widely scattered in the initial
 protoplanet region. 
We also find no difference in their eccentricities and inclinations, and
 those are $\simeq 0.1$.  
These eccentricities and inclinations are an order of magnitude larger
 than the proper eccentricities and inclinations of the present
 terrestrial planets.  
Some damping mechanism such as gravitational drag (dynamical friction)
 from a dissipating gas disk \citep{kji02} or a residual planetesimal disk
 \citep{acl99} is necessary after the giant impact stage.

\subsection{Statistics of Spin}


In 50 runs of the realistic and perfect accretion models, we have 128
 and 124 planets that experience at least one accretionary collision,
 respectively.   
The average values of each component of the spin angular velocity 
 {\boldmath $\omega$} of the planets and its dispersion $\sigma$ are
 summarized in Table~\ref{tab:spin} together with the Root-Mean-Square
 (RMS) spin angular velocity $\langle\omega^2\rangle^{1/2}$ and the spin
 anisotropy parameter
 $\beta=\langle\omega_z^2\rangle/\langle\omega^2\rangle$.  

\begin{figure}
\plottwo{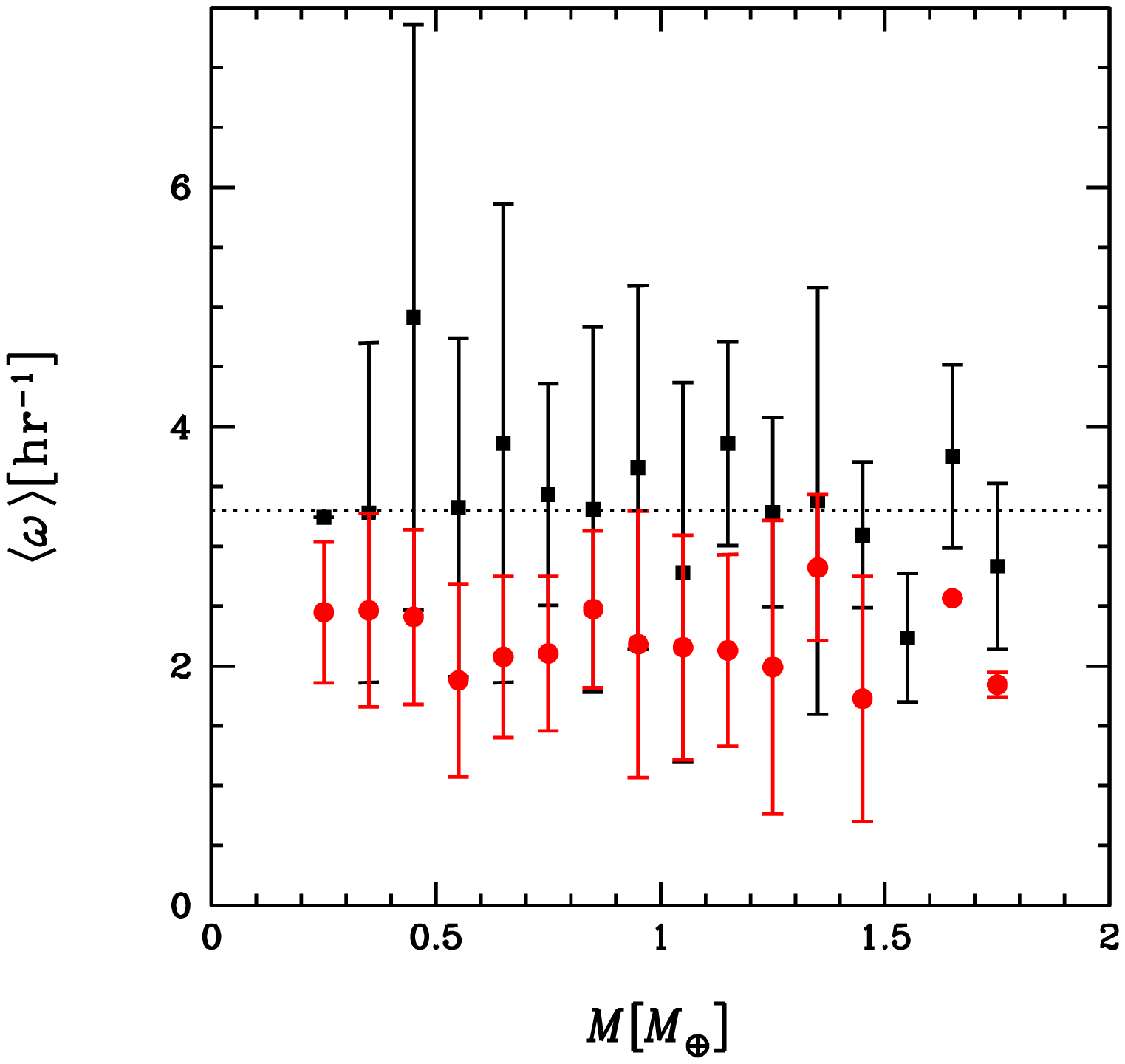}{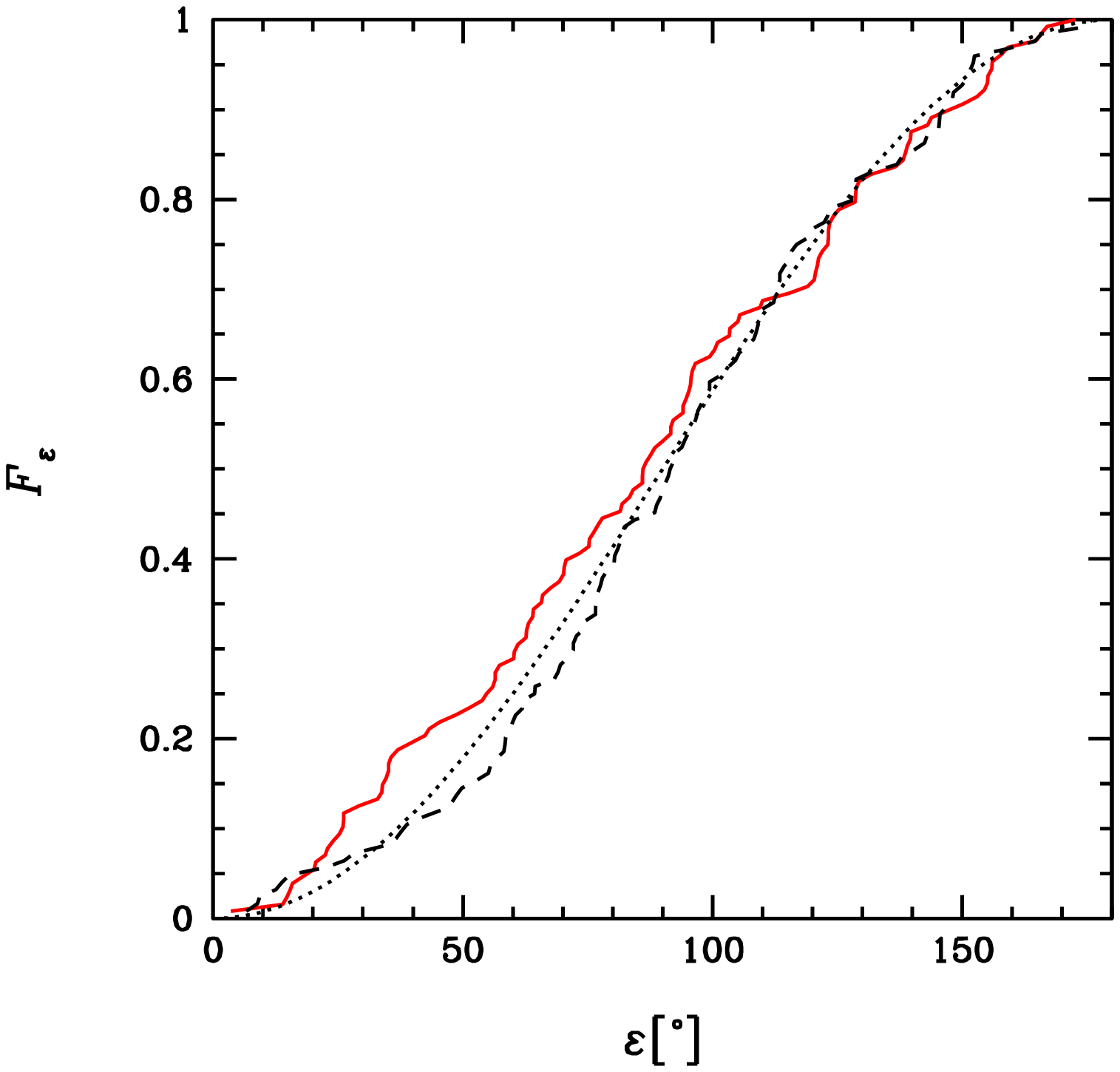}
\caption{
 \label{fig:m-w_ave_cmp}
Left: Average spin angular velocity of all planets formed in the 50 runs of
 the realistic (circle) and perfect (square) accretion models is plotted
 against their mass $M$ with mass bin of 0.1$M_\oplus$.
The error bars indicate 1-$\sigma$ and the dotted line shows
 $\omega\sub{cr}$. 
Right: 
Normalized cumulative distributions of $\varepsilon$ for the realistic
 (solid curve) and perfect (dashed curve) accretion models with an
 isotropic distribution (dotted curve).
}
\end{figure}

The spin angular velocity averaged in mass bins against mass is shown in 
 Figure~\ref{fig:m-w_ave_cmp}a.
We show clearly that the average angular velocity is almost
 independent of mass for both accretion models.
For the perfect accretion model, the average values are as high as the
 critical angular velocity $\omega\sub{cr}$.
These are natural outcomes for the giant impact stage under the
 assumption of perfect accretion \citep{ki07}.
The RMS spin angular velocity for the realistic accretion model is
 about 30\% smaller than that for the perfect accretion model.
This is because in the realistic accretion model, grazing and
 high-velocity collisions that have high angular momentum result in a 
 hit-and-run, while nearly head-on or low-velocity collisions that have
 small angular momentum lead to accretion. 
In other words, small angular momentum collisions are selective in
 accretion. 
Thus, the accretion inefficiency leads to slower planetary spin,
 compared with the perfect accretion model.
Indeed, the RMS spin angular velocity is almost consistent with the
 estimation by equation~(\ref{eq:w}) in
 section~\ref{sec:collision_parameter}.  
The RMS spin angular velocity slightly larger than the estimation
 based on a single dominant accretionary collision is due to the
 contribution of other non-dominant accretionary collisions. 

We confirm that each component of $\vec{\omega}$ follows a Gaussian
 distribution.
We perform a Kolmogorov-Smirnov (K-S) test to confirm the agreement
 with a Gaussian distribution.
For the distribution of each component, we obtain sufficiently high
 values of the K-S probability $Q\sub{KS}>0.5$ in both accretion models.  

In Figure~\ref{fig:m-w_ave_cmp}b, we show the obliquity distribution
 with an isotropic distribution, 
 $n d\varepsilon = (1/2)\sin \varepsilon d\varepsilon$.
We find that the obliquity ranges from 0$^\circ$ to 180$^\circ$ and
 follows an isotropic distribution 
 $n d\varepsilon = (1/2)\sin \varepsilon d\varepsilon$, which is
 consistent with \cite{acl99} and \cite{ki07}.
By a K-S test, we obtain high K-S probabilities of 0.3 and 0.9 for the
 realistic and perfect accretion models, respectively.   
This is also confirmed by the spin anisotropy parameter 
 $\beta \simeq 1/3$ in Table~\ref{tab:spin}.
The isotropic distribution of $\varepsilon$ is a natural outcome of
 giant impacts since the impacts are three-dimensional and add equally
 random contributions to each component of the spin angular momentum
 \citep{ki07}. 


For an Earth mass planet, the RMS spin angular velocity 2.35 hr$^{-1}$
 of the realistic accretion model corresponds to the spin angular
 momentum of $9.5\times 10^{41}$ g cm$^2$ s$^{-1}$, which is 2.7 times
 larger than the angular momentum of the Earth-Moon system. 
So the angular momentum of the Earth-Moon system is not a typical value
 of the realistic accretion model but it is reasonably within the
 distribution of $\omega$.

\section{Summary and Discussion}
\label{sec:summary}


We have investigated the basic dynamical properties of the terrestrial
 planets assembled by giant impacts of protoplanets by using $N$-body
 simulations.    
For the first time, we adopted the realistic accretion condition of
 protoplanets obtained by the SPH collision experiments.
The basic dynamical properties have been studied statistically with
 numbers of $N$-body simulations.
For the standard protoplanet system, the statistical properties
 of the planets obtained are the following:  
\begin{itemize}
 \item About half of collisions in the realistic accretion model do not
       lead to accretion.
       However, this accretion inefficiency barely lengthens the
       growth timescale of planets. 
 \item The numbers of planets are $\langle n\rangle \simeq 3{\rm -}4$ and 
       $\langle n_M\rangle \simeq \langle n_a\rangle \simeq2$. 
       The growth timescale is about $6{\rm -}7\times 10^7$ years.
       The masses of the largest and second-largest planets are
       $\langle M_1\rangle\simeq 1.2M_\oplus$ and 
       $\langle M_2\rangle\simeq 0.7M_\oplus$.  
       The largest planets tend to form around 
       $\langle a_1\rangle\simeq 0.8$AU, while $a_2$ is widely scattered
       in the initial protoplanet region.
       Their eccentricities and inclinations are $\simeq 0.1$.
       These results are independent of the accretion model.
 \item The RMS spin angular velocity for the realistic accretion model
       is about 30\% smaller than that for the perfect accretion model
       that is as large as the critical spin angular velocity for
       rotational instability.
       The spin angular velocity and obliquity of planets obey Gaussian
       and isotropic distributions, respectively, independently of the
       accretion model. 
\end{itemize}



We confirm that except for the magnitude of the spin angular velocity,
 the realistic accretion model gives the same results as the perfect
 accretion model.
This agreement justifies the use of the perfect accretion model to
 investigate the basic dynamical properties of planets except for the
 magnitude of the spin angular velocity.


In the present realistic accretion condition, we do not consider the
 effect of the spin on the accretion condition that would potentially
 change collisional dynamics. 
It is difficult to derive an accretion condition in terms of spin
 parameters by SPH collision simulations since the collision parameter
 space becomes huge and it is almost impossible to cover all parameter
 space. 
The fragmentation of planets is not taken into account, either.
Including the fragmentation may be able to further reduce the spin
 angular velocity of planets by producing unbound collisional fragments
 with high angular momentum. 
Furthermore, the collisional fragments can potentially alter orbital
 dynamics of planets through dynamical friction if their mass is large
 enough. 
The high eccentricities and inclinations of planets may be damped by
 dynamical friction from the collisional fragments. 
In order to take into account these effects and make the model of
 terrestrial planet formation more realistic, we plan to perform
 $N$-body simulations for orbital dynamics and SPH simulations 
 for collisional dynamics simultaneously in a consistent way in future
 work.

\medskip

We thank Shigeru Ida for his continuous encouragement.
This research was partially supported by MEXT (Ministry of
 Education, Culture, Sports, Science and Technology), Japan, the
 Grant-in-Aid for Scientific Research on Priority Areas, ``Development of
 Extra-Solar Planetary Science,'' and the Special Coordination Fund for
 Promoting Science and Technology, ``GRAPE-DR Project.''

\begin{deluxetable}{rrrrrrrrrrrrrrrrrr}
\tabletypesize{\scriptsize}
\tablewidth{0pt}
\rotate
\tablecaption{Basic Dynamical Properties of Final Planets\label{tab:number}} 
\tablehead{
 \colhead{accretion model} &
 \colhead{$\langle n\rangle$} &
 \colhead{$\langle n_M\rangle$} &
 \colhead{$\langle n_a\rangle$} &
 \colhead{$\langle f_a\rangle$} &
 \colhead{$\langle T\sub{grow}\rangle$($10^8$ yr)} &
 \colhead{$\langle n\sub{har}\rangle$} &
 \colhead{$\langle n\sub{acc}\rangle$} &
 \colhead{$\langle M_1\rangle$($M_\oplus$)} &
 \colhead{$\langle a_1\rangle({\rm AU})$} & 
 \colhead{$\langle e_1\rangle$} & 
 \colhead{$\langle i_1\rangle$} &
 \colhead{$\langle M_2\rangle$($M_\oplus$)} &
 \colhead{$\langle a_2\rangle({\rm AU})$} & 
 \colhead{$\langle e_2\rangle$} & 
 \colhead{$\langle i_2\rangle$}\\
} 
\startdata
realistic &
3.6$\pm$0.8 & 2.0$\pm$0.5 & 1.6$\pm$0.6 & 0.63$\pm$0.2 
& 0.73$\pm$0.74 & 11.8$\pm$7.7 & 12.4$\pm$0.8 & 
 1.18$\pm$0.23 & 0.79$\pm$0.27 & 0.12$\pm$0.07 & 0.07$\pm$0.05 &
 0.72$\pm$0.20 & 0.91$\pm$0.50 & 0.15$\pm$0.07 & 0.10$\pm$0.05 \\
perfect & 
3.1$\pm$0.6 & 2.0$\pm$0.6 & 1.7$\pm$0.6 & 0.77$\pm$0.2 
& 0.60$\pm$0.71 & \nodata & 12.9$\pm$0.6 &
 1.26$\pm$0.25 & 0.77$\pm$0.24 & 0.12$\pm$0.06 & 0.06$\pm$0.04 &
 0.72$\pm$0.23 & 0.97$\pm$0.52 & 0.14$\pm$0.08 & 0.09$\pm$0.06 
\enddata
\end{deluxetable}


\begin{deluxetable}{rrrrrrrrr}
\tablewidth{0pt}
\rotate
\tablecaption{Spin Parameters of Final Planets \label{tab:spin}}
\tablehead{
 \colhead{accretion model} &
$\langle \omega_x\rangle$(${\rm hr}^{-1}$) &
$\langle \omega_y\rangle$(${\rm hr}^{-1}$) &
$\langle \omega_z\rangle$(${\rm hr}^{-1}$) &
$\sigma_x$(${\rm hr}^{-1}$) &
$\sigma_y$(${\rm hr}^{-1}$) &
$\sigma_z$(${\rm hr}^{-1}$) &
$\langle \omega^2\rangle^{1/2}$(${\rm hr}^{-1}$) &
$\beta$
 } 
\startdata
realistic & 
 0.05 & 0.08 & 0.07 & 
 1.32 & 1.26 & 1.49 & 
 2.35 & 0.40 \\
perfect & 
 0.42 & -0.09 & -0.16 &
 2.15 & 2.17 & 2.03 & 
 3.69 & 0.31
\enddata 
\end{deluxetable}


\begin{thebibliography}{}


\bibitem[Agnor \&  Asphaug (2004)]{aa04}
Agnor, C., \& Asphaug, E. 2004, \apj, 613, L157

\bibitem[Agnor et al. (1999) Agnor, Canup, \& Levison]{acl99}
Agnor, C. B., Canup, R. M., \& Levison, H. F. 1999,
\icarus, 142, 219

\bibitem[Canup (2004)]{c04}
Canup, R. M. 2004, \icarus, 168, 433

\bibitem[Chambers (2001)]{c01}
Chambers, J. E. 2001, \icarus, 152, 205



\bibitem[Hayashi (1981)]{h81}
Hayashi, C. 1981, Prog. Theor. Phys. Suppl., 70, 35

\bibitem[Ida \& Makino (1992)]{im92}
Ida, S., \& Makino, J. 1992, \icarus, 96, 107

\bibitem[Kokubo \& Ida (1998)]{ki98}
Kokubo, E., \& Ida, S. 1998, \icarus, 131, 171

\bibitem[Kokubo \& Ida (2000)]{ki00}
Kokubo, E., \& Ida, S., 2000, \icarus, 143, 15

\bibitem[Kokubo \& Ida (2002)]{ki02}
Kokubo, E., \& Ida, S., 2002, \apj, 581, 666

\bibitem[Kokubo et al. (2006) Kokubo, Kominami, \& Ida]{kki06}
Kokubo, E., Kominami, J., \& Ida, S., 2006, \apj, 642, 1131

\bibitem[Kokubo \& Ida (2007)]{ki07}
Kokubo, E., \& Ida, S., 2007, \apj, 671, 2082

\bibitem[Kokubo \& Makino (2004)]{km04}
Kokubo, E., \& Makino, J. 2004, \pasj, 56, 861

\bibitem[Kokubo et al. (1998) Kokubo, Yoshinaga, \& Makino]{kym98}
Kokubo, E., Yoshinaga, K., \& Makino, J. 1998, \mnras, 297, 1067

\bibitem[Kominami \& Ida (2002)]{kji02}
Kominami, J., \& Ida, S. 2002, \icarus, 157, 43

\bibitem[Makino (1991)]{m91}
Makino, J. 1991, \pasj, 43, 859

\bibitem[Melosh (1989)]{m89}
Melosh, H. J. 1989, Impact Cratering: A Geologic Process (New York: Oxford 
Univ. Press)

\bibitem[Monaghan (1992)]{m92}
Monaghan, J. J. 1992, \araa, 30, 543

\bibitem[Nitadori et al. (2006) Nitadori, Makino, \& Hut]{nmh06}
Nitadori, K., Makino, J., \& Hut, P. 2006, New Astronomy, 12, 169.


\bibitem[Wetherill (1985)]{w85}
Wetherill, G. W. 1985, Science, 228, 877


\end{thebibliography}
\end{document}